\documentclass[prd,12pt,aps,superscriptaddress,double-spaced,floatfix,showkeys,showpacs,twocolumn]{revtex4-1}
\usepackage{graphicx,amsmath,bbm,bm,array,subfigure}
\usepackage{appendix}
\usepackage{lipsum}
\usepackage[linktocpage=true,colorlinks=true,linkcolor=blue,citecolor=blue]{hyperref}
\usepackage{MnSymbol}

\begin{document}

\title { Effects of  hadronic repulsive interactions on the  fluctuations of conserved charges}

\author{Somenath Pal}
\email{somenathpal1@gmail.com}
\affiliation{Department of Physics, University of Calcutta, 92, A.P.C. Road, Kolkata-700009, India}

\author{Guruprasad Kadam }
\email{guruprasadkadam18@gmail.com}
\affiliation{Department of Physics,
	Shivaji University, Kolhapur,
	Maharashtra-416004, India}
\author{Hiranmaya Mishra}
\email{hm@prl.res.in}
\affiliation{Theory Division, Physical Research Laboratory,
 Navarangpura, Ahmedabad - 380 009, India}
  \author{Abhijit Bhattacharyya}
 \email{abhattacharyyacu@gmail.com}
 \affiliation{Department of Physics, University of Calcutta, 92, A.P.C. Road, Kolkata-700009, India}
\date{\today} 

\def\be{\begin{equation}}
\def\ee{\end{equation}}
\def\bearr{\begin{eqnarray}}
\def\eearr{\end{eqnarray}}
\def\zbf#1{{\bf {#1}}}
\def\bfm#1{\mbox{\boldmath $#1$}}
\def\hf{\frac{1}{2}}
\def\sl{\hspace{-0.15cm}/}
\def\omit#1{_{\!\rlap{$\scriptscriptstyle \backslash$}
{\scriptscriptstyle #1}}}
\def\vec#1{\mathchoice
        {\mbox{\boldmath $#1$}}
        {\mbox{\boldmath $#1$}}
        {\mbox{\boldmath $\scriptstyle #1$}}
        {\mbox{\boldmath $\scriptscriptstyle #1$}}
}

\begin{abstract}
We investigate the effect of repulsive interaction between hadrons on the fluctuations  of the 
 conserved charges.  We calculate the baryon,  the electric charge and the strangeness 
susceptibilities within the ambit of  hadron resonance gas model extended to include the 
short range repulsive interactions. 
The repulsive interactions are included through  a mean-field approach where the single particle
 energy gets modified
 due to mean field interactions between hadrons proportional to the number density of hadrons. 
We assume different mean-field interactions for mesons and baryons. It is shown that the 
repulsive interactions play  a very crucial role to describe hadronic matter near transition temperature. We also show 
that in order to consistently describe higher order conserved charge fluctuations mesonic 
repulsive interactions cannot be 
neglected. Further, we demonstrate that the repulsive interaction of baryons are 
essential to describe 
the lattice simulation results at finite baryon chemical potential for higher order fluctuations.
\end{abstract}


\maketitle

\section{Introduction}
Studies of strongly interacting matter at high temperatures and/or densities have been a vibrant area of research for several decades. 
At such high temperatures and/or densities the effective degrees of freedom of the strongly interacting matter are colored quarks and gluons 
whereas at low temperatures and/or densities these are colorless hadrons.  There are several ongoing and up-coming
experiments with ultra-relativistic heavy ion collision which are
recreating such phases of strongly interacting matter. The experimental
programs, Large Hadron Collider (LHC)  at Geneva and Relativistic Heavy Ion Collider (RHIC) at Brookhaven, 
have already provided a plethora of data. The 
upcoming facilities like Facility for Antiproton and Ion Research (FAIR) and Nuclotron-based Ion Collider fAcility (NICA) 
 are expected to shed more light in this area of research. 

 One of the main objectives of these explorations is to understand the 
 phase diagram of strongly interacting matter.   At high temperature and 
at vanishing or small values of baryon chemical potential, the transition between the
hadronic matter and the quark-gluon matter is established  to be a crossover~\cite{crossover1,crossover2}. On the other hand,  at low temperature and high baryon densities it is generally expected that such a transition is possibly a first order phase 
transition~\cite{1st_order1, 1st_order2, 1st_order3, 1st_order4, 1st_order5,1st_order6}. Therefore, in the phase diagram of strong interaction in the temperature and baryon chemical potential plane, one expects a first order line ending in a critical point ~\cite{CEP1, CEP2, CEP3, CEP4}. The search for this elusive critical end 
point (CEP) of  strong interaction has become an extremely active field of research since
last few years although its existence as a fundamental property of strong interaction still 
remains to be confirmed experimentally. The CEP  is typically characterised by large fluctuations 
of the static thermodynamic quantities due to large correlation length. One of the crucial realisation 
 in this context 
has been the fact that the measurement of the moments of the conserved quantities
 viz., baryon number, charge and strangeness,
 can be accessible 
in heavy-ion collisions experiments and can be linked to
susceptibilities that can be computed in QCD based calculations.
 In order to confirm, whether the thermalised system, produced in heavy-ion collisions, 
has passed through the quark-hadron phase transition, it is necessary to make  theoretical calculations of the  fluctuations of
 conserved charges from both sides of the phase transition line.

Though quantum chromodynamics (QCD) is the theory of strong interactions, the traditional perturbative 
methods of field theory cannot 
be applied in the case of temperature/density of interest. 
Lattice QCD (LQCD) has been the principle tool to understand the
equlibrium  phase structure of strongly interacting matter and the 
fluctuations of conserved charges have been extensively studied~\cite{crossover1,LQCD1, LQCD2, LQCD3, LQCD4, LQCD5, LQCD6,
LQCD7, LQCD8, LQCD9, LQCD10, LQCD11, LQCD12, LQCD13, LQCD14, LQCD15}. Despite its success at zero baryon density, LQCD
 calculations have limited applicability at finite baryon density. Hence in the finite density region of QCD phase diagram 
one needs  to take resort to effective models of QCD which preserve some essential properties of QCD at a given energy scale. 
For instance, Polyakov-Nambu-Jona-Lasinio (PNJL) model~\cite{PNJL1,PNJL2,PNJL3,PNJL4,PNJL5,PNJL6,PNJL7,PNJL8,PNJL9,PNJL10,PNJL11,PNJL12,
PNJL13,PNJL14,PNJL15,PNJL16,PNJL17},
Hadron Resonance Gas (HRG) Model~\cite{HRG1,HRG2,HRG3,HRG4,HRG5,HRG6,HRG7,HRG8,HRG9,HRG10,HRG11}, Polyakov-Quark-Meson (PQM) 
model~\cite{PQM1,PQM2,PQM3,PQM4}, Chiral Perturbation Theory~\cite{chiral1}, etc. have been rather
 successful in describing different parts of the finite density region of the phase diagram. These models have also been found 
 to reproduce the lattice data at zero density quite successfully~\cite{PNJL11,PNJL12}.

   Hadron resonance gas model is a low temperature statistical thermal model describing hadronic phase of QCD. This model 
is based on S-matrix formulation of statistical mechanics~\cite{Dashen:1969ep}. In the relativistic virial expansion of 
the partition function the interactions are manifested in the form of  phase shifts in the two particle scattering. 
If such scattering occurs through exchange of a  narrow resonance state then the interacting partition function 
just becomes  a non-interacting partition function with the additional contribution from the exchanged 
resonance~\cite{Dashen:1974yy,Welke:1990za,Venugopalan:1992hy}. 
Indeed, it turns out that the non resonant terms cancel out in the estimation of certain thermodynamic quantities.
 Thus, in the HRG model the thermodynamic quantities 
of low temperature hadronic matter  can be obtained from the partition function which contains all relevant degrees of freedom 
of the confined QCD phase and  implicitly includes the (attractive) interactions that result in resonance formation. 
     
   Despite its success, it was soon realized that non-interacting HRG model is not sufficient to describe hadronic matter,
particularly, near quark-hadron transition temperature, $T_c$. As temperature increases gas density increases and at 
high temperature the assumption of dilute gas approximation, a principal assumption of HRG model, need not be valid. 
The hadronic repulsive interaction becomes increasingly important as one approaches the critical temperature.
The validity of HRG model can only be checked by confronting its equation of state (EoS) with LQCD simulations. Various studies
 have confronted ideal HRG EoS with lattice results and found reasonable agreement all the way upto $T_c$ except 
for interaction measure~\cite{Karsch:2003vd}. Later studies found good agreement with the lattice data for trace anomaly 
as well when continuous Hagedorn states are included in HRG model~\cite{NoronhaHostler:2008ju,Kadam:2014cua}. However, 
 recent studies have shown that commonly performed  comparisons of ideal HRG model with LQCD simulations and heavy-ion data may 
lead to misconceptions which might further render wrong conclusions and it is necessary to take into account short range 
repulsive interactions  among the hadrons as they play  a crucial role in the thermodynamics 
of hadron gas~\cite{Vovchenko:2016rkn}.
 
 In  the present investigation, we would like to estimate conserved charge fluctuations within the ambit 
of an interacting HRG model that includes the repulsive interactions.  
Equilibrium fluctuations, of conserved charges, has been studied using the ideal HRG 
model~\cite{Karsch:2010ck, Garg:2013ata, Samanta:2018ufe,Mishra:2016tne} 
(See Ref.~\cite{Asakawa:2015ybt} 
for a review). This analysis showed that the lower order cumulants 
are reasonably well reproduced but higher order cumulants show significant deviation.
One possible way to include such repulsive interaction is through the the van der Walls excluded volume approach
as in Ref.s~\cite{rischke,singh}. Another approach has been to include repulsive interaction through a
repulsive density dependent mean 
field ~\cite{Kapusta:1982qd,Olive:1980dy,Huovinen:2017ogf,Kadam:2019peo}. 
Such a mean field HRG (MFHRG) model has been used to compute
fluctuations in the net baryon number and the strangeness-baryon correlation at vanishing 
chemical potentials~\cite{Huovinen:2017ogf}.
This was also considered for
hadron thermodynamics and transport properties~\cite{Kadam:2019peo} of hadronic matter.

Specifically, we wish to estimate in the present work, the second and the fourth 
order fluctuations for baryon number ($B$), electric charge ($Q$) and strangeness ($S$) quantum numbers 
using mean-field HRG model (MFHRG).
 We  also estimate the ratios as well as the differences of second and fourth order fluctuations. 
While the ratios of fluctuations are important so as to remove the effect of system volume, differences of fluctuations 
remove the effect of mass spectrum included in HRG model and one can distill the effect of only repulsive interactions 
on the fluctuations. It may further be noted that differences of susceptibilities are also indicators of 
deconfinement phase transition~\cite{Bazavov:2013dta}. We estimate these susceptibilities both at vanishing as well as finite baryon densities.

We organize the paper as follows. In Sec.~\ref{secII} we discuss the thermodynamics of relativistic mean-field hadron 
resonance gas model and define the susceptibilities. Results of cumulants are discussed in Sec.~\ref{secIII}.  Finally,
 In Sec.~\ref{secIV} we  summarize our findings and present our conclusions.

\section{Hadron resonance gas model with a repulsive mean field interaction}
\label{secII}

Thermodynamic properties of ideal hadron resonance gas model can be derived from the grand canonical partition function given by
\begin{eqnarray}
\text{ln}\mathcal{Z}(&T,\mu_B,\mu_Q,\mu_S, V)= 
\sum_{ \text{mesons}}\text{ln}\mathcal{Z_i}(T,\mu_Q,\mu_{S},V) \nonumber \\
&+\sum_{ \text{baryons}}
\text{ln}\mathcal{Z_a}(T,\mu_B,\mu_Q,\mu_{S},V)
\end{eqnarray}
where $\mu_B, \mu_Q$ and $\mu_S$ are the chemical potentials corresponding to baryon number, electric charge and strangeness respectively. The pressure can be obtained from the partition function as
\bearr
P(T,\mu_B,\mu_Q,\mu_S,V)= \nonumber \\
\lim_{V\rightarrow\infty}\frac{ T}{V}\  \text{ln}\mathcal{Z}(T,\mu_B,\mu_Q,\mu_S,V)
\eearr

Partition function of $i^{\text{th}}$ hadronic species is 

\be
\text{ln} \mathcal {Z_a}(T,V,{\mu_a})=\pm V\int d\Gamma_{a}\: \text{ln}[1\pm e^{-\frac{(E_a-{c_i}
{\mu_i})}{T}}]
\ee
where, for the i-th species of hadrons,  $c_{i}\equiv(B_i,Q_i,S_i)$ corresponds to the respective
conserved charges and ${\mu_i}\equiv (\mu_B, \mu_Q, \mu_S)$ is the corresponding chemical potential. Also,
 $d\Gamma_{a}\equiv\frac{g_{a}d^{3}p}{(2\pi)^3}$ with $g_a$ being the spin degeneracy factor and 
 $E_{a}=\sqrt{p^2+m_a^2}$ is the relativistic  single particle energy with mass $m_a$. 
 Upper (lower) sign corresponds to fermions (bosons). 

Ideal HRG model can be extended by including short range repulsive interactions between hadrons. These 
repulsive interactions can be treated in mean field approach where the single particle energies 
 $\epsilon_a$ get shifted by the mean field repulsive interaction 
 as ~\cite{ Kapusta:1982qd,Olive:1980dy}
\be
\varepsilon_{a}=\sqrt{p^2+m_{a}^2}+U(n)=E_{a}+U(n)
\label{dispersion}
\ee
where $n$ is the total hadron number density. The potential energy $U$ represents repulsive interaction between hadrons 
and it is taken to be function of total hadron density $n$. For an arbitrary inter-hadron potential $V({\bf{r}})$, 
the potential energy $U$  is given by

\be
U(n)=Kn
\label{poten}
\ee
where $K$ is a  phenomenological parameter given by the spatial integration of the inter hadron potential $V({\bf r})$.

In the present investigation, we  assume different repulsive interaction parameter for baryons  and mesons. 
We denote the mean field parameter for baryons ($B$) and anti-baryons ($\bar{B}$) by $K_B$, while for mesons 
we denote it by  $K_M$. Thus,  for baryons (antibaryons)
\be
U(n_{B\{\bar{B}\}})=K_Bn_{B\{\bar{B}\}}
\label{potenbar}
\ee

and for mesons
\be
U(n_M)=K_Mn_M
\label{potenmes}
\ee

The total hadron number density is

\be
n(T,\mu)=\sum_{a}n_{a}=n_B+n_{\bar{B}}+n_M
\ee

where $n_{a}$ is the number density of a$^\text{th}$ hadronic species. Also, $n_B$, $n_{\bar{B}}$  and $n_M$ are total baryon, anti-baryon and meson number densities respectively. For baryons,

\be
n_{{B}}=\sum_{a\in B}\int d\Gamma_{a}\:\frac{1}{e^{\frac{(E_{a}-{\mu_{\text{eff},B}})}{T}}+1}
\label{numdenbaryon}
\ee
where $\mu_{\text{eff},{B}}=c_i\mu_i-K_Bn_{B}$ and $c_i =(B_i,Q_i,S_i), {\mu_i}=(\mu_B,\mu_Q ,\mu_S)$.
 The sum is over all the baryons. Similarly, the number density of antibaryons is

\be
n_{{\bar{B}}}=\sum_{a\in \bar{B}}\int d\Gamma_{a}\:\frac{1}{e^{\frac{(E_{a}-{\mu_{\text{eff},\bar{B}}})}{T}}+1}
\label{numdenantibaryon}
\ee
where $\mu_{\text{eff},{\bar B}}=\bar{c_i}{\mu_i}-K_Bn_{\bar{B}}$. Note that repulsive mean-field parameter is same for baryons as well as anti-baryons.
 For mesons,

\be
n_{{M}}=\sum_{a\in M} \int d\Gamma_{a}\:\frac{1}{e^{\frac{(E_{a}-{\mu_{\text{eff},M}})}{T}}-1}
\label{numdenmeson}
\ee
where $\mu_{\text{eff},{M}}={c_i}\mu_i-K_Mn_{M}$ and the sum is over all the mesons. 

Eqs. (\ref{numdenbaryon})-(\ref{numdenmeson})  are actually self consistent equations for number densities which can be solved numerically.
The expressions for the pressures  of baryons and mesons are then given respectively as
\bearr
&&P_{B\{\bar{B}\}}(T,\mu)=T\sum \limits_{a\in B \{\bar B \}} \int d\Gamma_{a} \nonumber\\ 
&& \text{ln}\bigg[1+ e^{-(\frac{E_a-\mu_{\text{eff}}\{\bar{\mu}_{\text{eff}}\}}{T})}\bigg] \nonumber\\
&&-\phi_{B\{\bar{B}\}}(n_{B\{\bar{B}\}})
\label{pbbar}
\eearr
\bearr
&P_M(T,\mu)=-T \sum_{a\in M} \int d\Gamma_{a}\text{ln}\bigg[1- {e^{\frac{(E_{a}-{\mu_{\text{eff},M}})}{T}}}\bigg] \nonumber\\
&-\phi_M(n_M)
\eearr
where,

\be
\phi_B(n_{B\{\bar{B}\}})=-\frac{1}{2}K_Bn_{B\{\bar{B}\}}^2
\ee
and
\be
\phi_M(n_M)=-\frac{1}{2}K_Mn_M^2
\ee

Thermodynamic quantities can be readily calculated by taking appropriate derivative of the partition function 
or equivalently of the pressure.

The nth-order susceptibility is defined as 
\begin{equation}
\chi_{i}^{n}= T^n\frac{\partial^n(P(T,\mu_i)/T^4 )}{\partial (\frac{\mu_{i}}{T})^n}
\label{chin}
\end{equation}
where $\mu_i$ is the chemical potential for conserved charge $c_i$. In this work we will take $i$ to be baryon number ($B$),
 electric charge ($Q$) and strangeness ($S$). The $n=1,2,3,4$ susceptibilities are related to the mean, variance, skewness and kurtosis of the
distribution of the conserved charges. 

Before proceeding further, let us discuss some approximate  expressions for the pressure and the number densities 
and hence on the susceptibilities which is useful to analyse the behaviour of susceptibilities with temperature and/or 
chemical potentials.
One can expand the logarithm in the expression for (non-interacting HRG) pressure in powers of  fugacity so that 
the baryonic pressure in 
Eq.(\ref{pbbar}) can be written as, (with $\beta=T^{-1}$)
\bearr
\frac{{P}_{B\{\bar B\}}}{T^4}&=&\sum\limits_{a\in B\{\bar B\}}\frac{g_a}{2\pi^2}(\beta m)^2\sum\limits_{l=1}^{\infty}
(-1)^{l+1} \:l^{-2} \nonumber\\
&\times& K_2(\beta l m_a)z^{l} \nonumber\\
&+&\frac{K_BT^2}{2}\left(\frac{n_{B\{\bar B\}}}{T^3}\right)^2
\label{bessell}
\eearr
In the above, we have defined the fugacity as $z=\exp ({\beta\mu_{eff}})$ and $ K_2$ is the Bessel function. It can be easily 
shown that as long
as $\beta(m_a-\mu_{eff})\gtrsim 1$, the contribution to the pressure $P^{id}_{B\{\bar B\}}$
 can be approximated by the leading term i.e. $l=1$ in the summation which, in fact, corresponds to Boltzmann approximation. In this limit,
the pressure from the baryons become
\bearr
&\frac{P_{B\{\bar B\}}}{T^4}
=\sum_{a\in B}\frac{g_a}{2\pi^2}(\beta m_a)^2K_2(\beta m_a)\nonumber\\
&\times \exp(\beta\mu_{eff}^a)+\frac{K_BT^2}{2}\left(\frac{n_{B\{\bar B\}}}{T^3}\right)^2
\eearr

In a similar approximation, the number density for baryons can be written as
\be
\frac{n_{B}}{T^3}=\sum_{a\in{B}}\frac{g_a}{2\pi^2}(\beta m)^2K_2(\beta m_a)e^{\beta\mu_{eff}^a}
\label{approxnb}
\ee

Thus, the interacting gas pressure in the Boltzmann approximation can be written as
\be
P_{B\{\bar B\}}=T n_{B\{\bar B\}}+\frac{K_B}{2} n_{B\{\bar B\}}^2
\ee



One can further approximate for the number densities given in Eq.(\ref{approxnb})
 by noting that, for temperatures below the
QCD transition temperatures such that $n_B$ $(n_{\bar B})$ are small, one
can expand the exponential  $\exp(\mu_{eff})\simeq \exp(c_i\mu_i)(1-\beta K_B n_B)$. This leads to
$n_B=n_B^{id}/(1+n_B^{id})$. Here $n_B^{id}$ is the number density without any repulsive interaction (Eq.(\ref{approxnbid})) i.e.
\be
\frac{n_{B}^{id}}{T^3}=\sum_{a\in{B}}\frac{g_a}{2\pi^2}(\beta m)^2K_2(\beta m_a)\exp(\beta c_a^i\mu_i)
\label{approxnbid}
\ee

.
The pressure due to baryons can then be approximated as
\be
P_B=Tn_B^{id}-\frac{K_B}{2}{(n_B^{id})^2}
\ee
A similar expression can be obtained for anti-baryon pressure. As may be noted, the effect of the density dependent repulsive 
interaction essentially lies in 
reducing the pressure at finite densities.

The total pressure from baryons and antibaryons can then be written as

\begin{eqnarray}
&&\frac{P_B+P_{\bar B}}{T^4}\simeq \sum_{a\in B}F_a(\beta m_a)\cosh(\beta c_a^i\mu_i)\nonumber\\
&-&\frac{K_B}{2}\sum_aG_a(\beta m_a,\beta\mu_Q,\beta\mu_s)e^{2\beta\mu_B}\nonumber\\
&-&\frac{K_B}{2}\sum_aG_a(\beta m_a,-\beta\mu_Q,-\beta\mu_s)e^{-2\beta\mu_B}.
\label{pbapprox}
\end{eqnarray}


Here we have defined the  chemical potential independent function $F_a(\beta m_a)$ as
\be
F_a(\beta m_a)=\frac{g_a}{\pi^2}(\beta m_a)^2 K_2(\beta m_a)
\label{Fa}
\ee
and, the baryon chemical potential independent function
\bearr
&&G_a(\beta m_a,\beta\mu_Q,\beta\mu_s)=\frac{g_a}{2\pi^2}(\beta m_a)^2 K_2(\beta m_a)\nonumber\\
&\times &\exp(Q_a\beta\mu_Q+S_a\beta\mu_s)
\label{Ga}
\eearr

%

In a similar manner, for the mesons, in the Boltzmann approximation, and with $\beta K_m n_m\le 1$, we have

\be
\frac{P_M}{T^4}\simeq \sum_{a\in M}\frac{n_a^{id}}{T^3}-\frac{1}{2} (K_MT^2)\bigg(\frac{n_a^{id}}{T^3}\bigg)^2
\label{pmapprox}
\ee
with
\be
n_a^{id}=\frac{g_a}{2\pi^2}K_2(\beta m_a) \text{exp}(\beta \mu_a)
\ee

The total pressure $P=P_B+P_{\bar B}+P_M$ is thus given approximately by the sum of Eqs.(\ref{pbapprox}) and (\ref{pmapprox}).
These approximate expressions for pressure have interesting consequences. Firstly, in the context of baryonic susceptibilities, 
the odd order susceptibilities will be  small for small chemical potential
and will vanish for zero baryonic chemical potential. Further, for the even order baryonic susceptibilities, 
e.g. $\chi^4_B$ and $\chi^2_B$ will be identical but for the repulsive interaction term. Indeed, 
the difference between these is given approximately as
\be
\chi^2_B-\chi^4_B\simeq 
\frac{K_B}{2} (1-(2\beta\mu_B)^2\chi_2^B(\beta\mu_B)
\ee
Thus, while  the difference between the higher(even) order and lower(even) order baryonic 
susceptibilities will vanish for
ideal HRG, it will not vanish when there is mean field repulsive terms. In the following we shall
discuss the results where we take the actual Fermir-Dirac or Bose-Einstein statistics for the hadrons
and solve for the self consistent equations for the number densities 
for the estimation of susceptibilities. However, as we shall observe, the above assertions made with the approximate
expressions for the pressure remain valid.

\section{Results and discussion}
\label{secIII}

In this section we are going to discuss the results of susceptibilities of different conserved quantities 
calculated from the MFHRG model. We will compare our results with those 
obtained from LQCD; namely with Ref.~\cite{LQCD16} for vanishing chemical potential and with Ref.~\cite{LQCD20}
for non-vanishing chemical potentials. Fluctuations of conserved charges like net baryon number, electric charge, 
strangeness are useful indicators of thermalisation and hadronization of matter produced in 
ultra relativistic heavy ion collision~\cite{PNJL11,PNJL12,Jeon,Koch,fluc1,fluc2,fluc3,fluc4,fluc5}. 
Large fluctuations in various thermodynamic quantities are important signatures of existence of Critical End Point (CEP) 
in the phase diagram.
To estimate the different thermodynamic quantities,
we have taken all the hadrons and resonances up to 3 GeV listed in particle data review~\cite{Tanabashi:2018oca}. The only parameters in our model are $K_M$ and $K_B$ as mentioned 
in the previous section. We choose three different representative values for meson mean field parameter,
 $viz.$, $K_M=0, 0.1$ and $0.15$ GeV$\cdot$fm$^{3}$,  while we fix baryon mean-field parameter 
$K_B=0.45$ GeV$\cdot$fm$^{3}$~\cite{Huovinen:2017ogf,Kadam:2019peo}.
 \begin{figure}[h]
	\vspace{-0.4cm}
	\begin{center}
		\begin{tabular}{c c}
			\includegraphics[width=8cm,height=8cm]{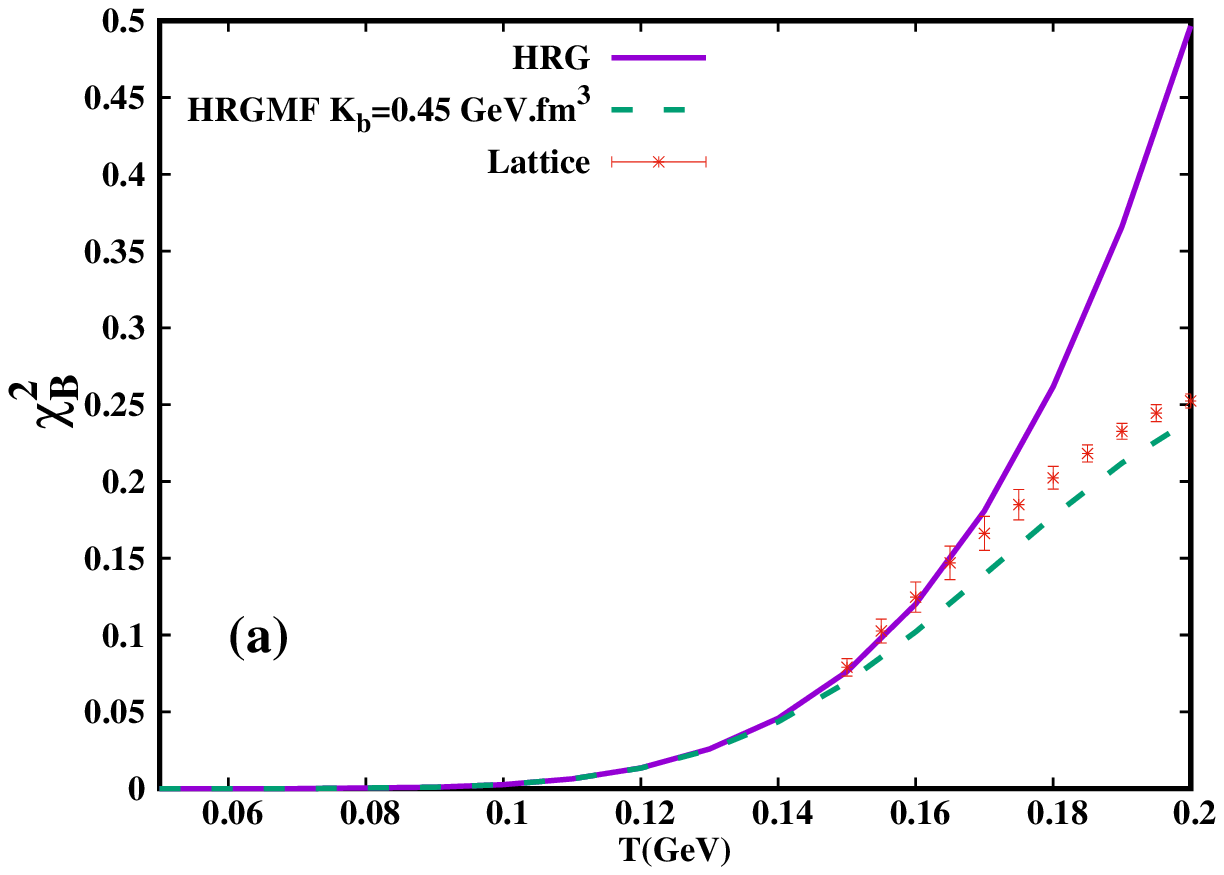} \\
			\includegraphics[width=8cm,height=8cm]{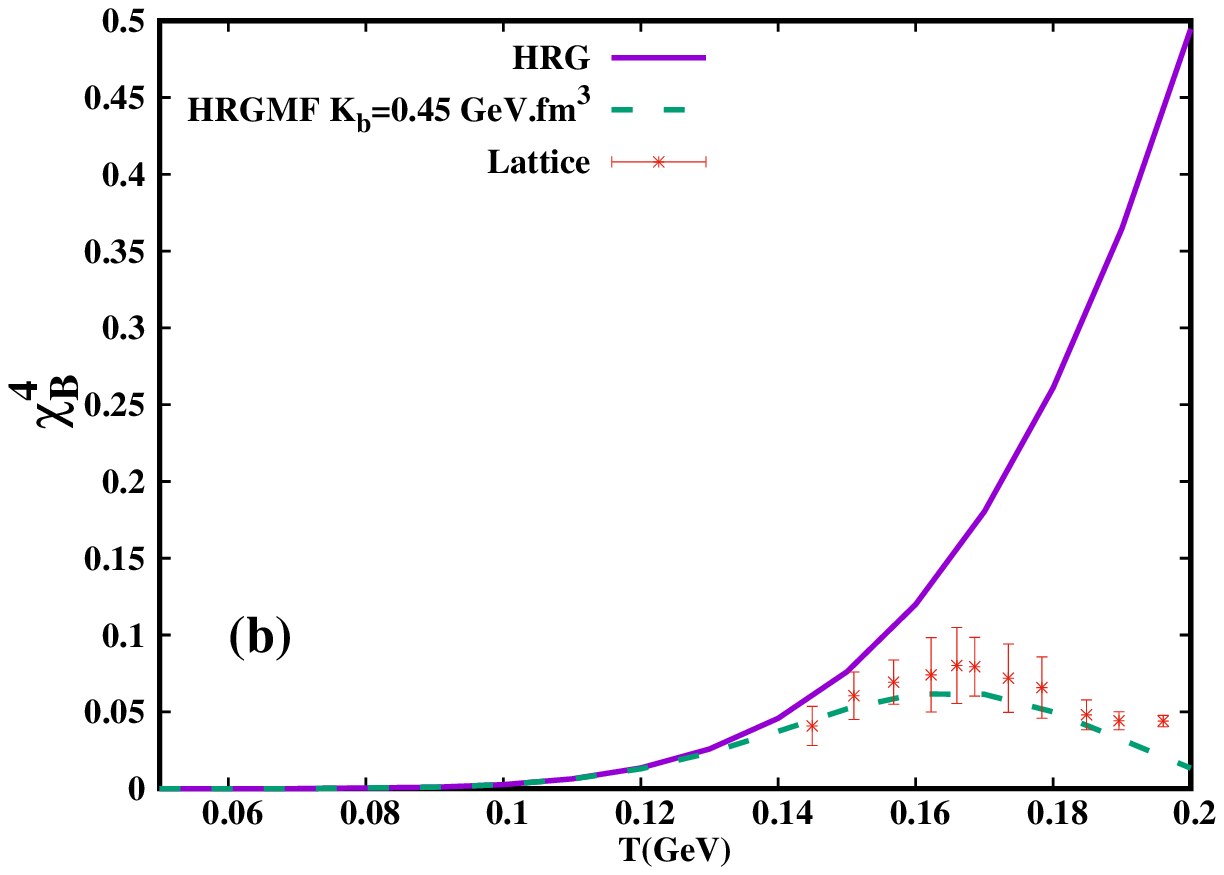}
		\end{tabular}
		\caption{(Color Online) Baryon number susceptibilities of different orders.  This result is independent of $K_m$. The lattice data is taken from Ref.~\cite{LQCD16}.} 
		\label{chiB}
	\end{center}
\end{figure}

In Fig.\ref{chiB} we plot the  2nd and 4th order baryon number susceptibilities. In the absence of 
any repulsion, the susceptibilities, calculated in the HRG model, increase monotonically with temperature. However, 
as  the interaction, among the baryons, is switched on through $K_B$, the susceptibilities (especially $\chi_B^4$) 
show non-monotonic behaviour. We note that HRG model reproduces the LQCD results up to a temperature 
of  $T=160$ MeV after which it deviates. On the other hand, MFHRG provides a very good qualitative and quantitative description. 
The broad bump in $\chi_B^4$ (Fig.\ref{chiB}(b)), as obtained in the lattice simulations, 
is very well reproduced in MFHRG model. The repulsive interaction reduces the  baryonic susceptibility results for
higher temperatures as may be expected from the approximate expressions for pressure in Eq.(\ref{pbapprox}). This also
explains that the reduction is more for the  higher order susceptibility.
Note that mesons do not contribute to $\chi_{B}$ and hence the results are independent of the value of $K_m$. 
 Since all the baryons have baryon number one, the numerical values of $\chi_B^2$ and $\chi_B^4$ are same for HRG.
 However, this is not true for the interacting (MFHRG) scenario. Agreement of MFHRG  with LQCD emphasises the 
role of repulsive interaction in the thermodynamics of hadron gas especially at higher temperature. 
Recently similar studies have observed that if we switch-off the repulsive interactions between mesons 
and include van-der Waals parameters characterising repulsive and attractive interaction between baryons, 
resulting model turns out to be in better agreement with the lattice data~\cite{Vovchenko:2016rkn}.  Thus, ideal HRG is insufficient to describe higher order susceptibilities and the agreement with the results from mean-field HRG, which takes into account repulsive interactions, indicate that these interactions cannot be neglected in the studies which are being carried out to probe quark-hadron phase transition as well as QCD critical point. 

 \begin{figure}[h]
	\vspace{-0.4cm}
	\begin{center}
		\begin{tabular}{c c}
			\includegraphics[width=8cm,height=8cm]{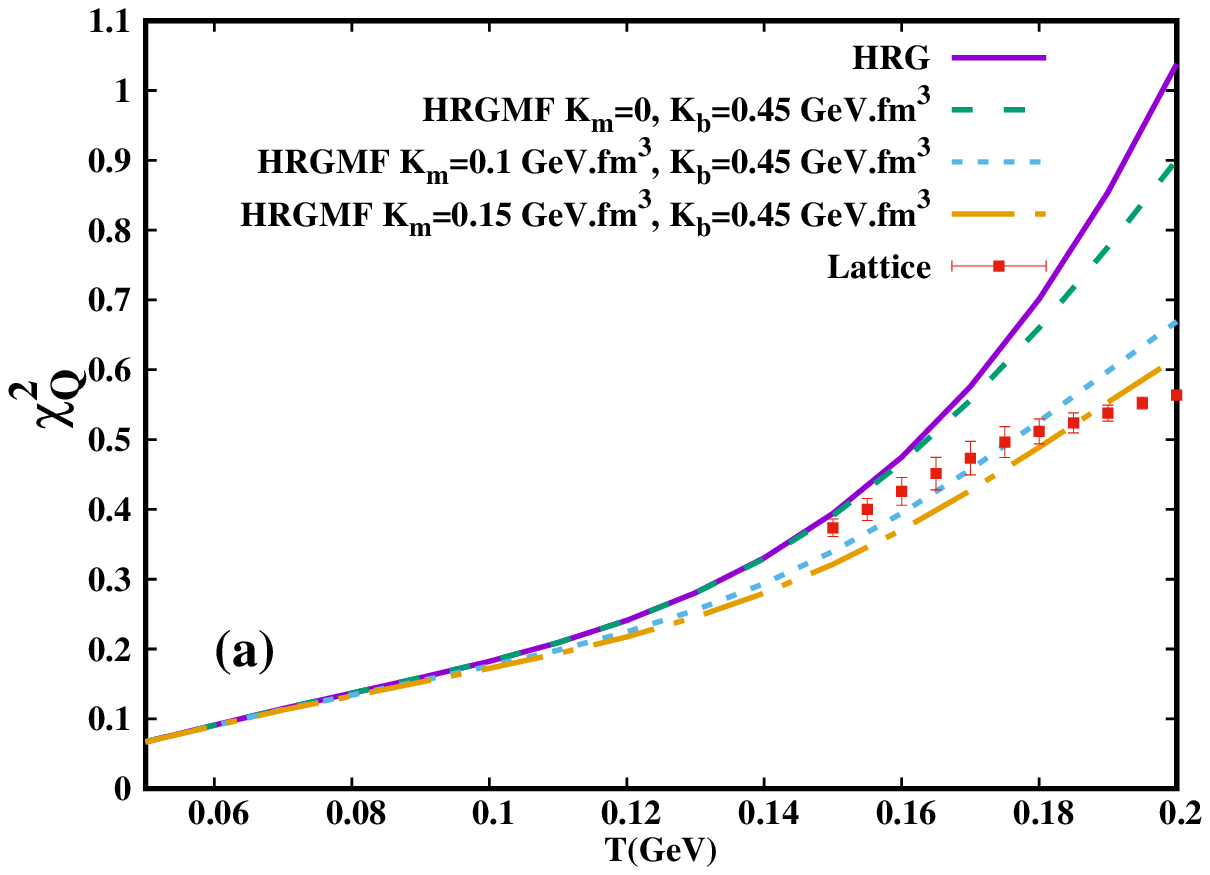} \\
			\includegraphics[width=8cm,height=8cm]{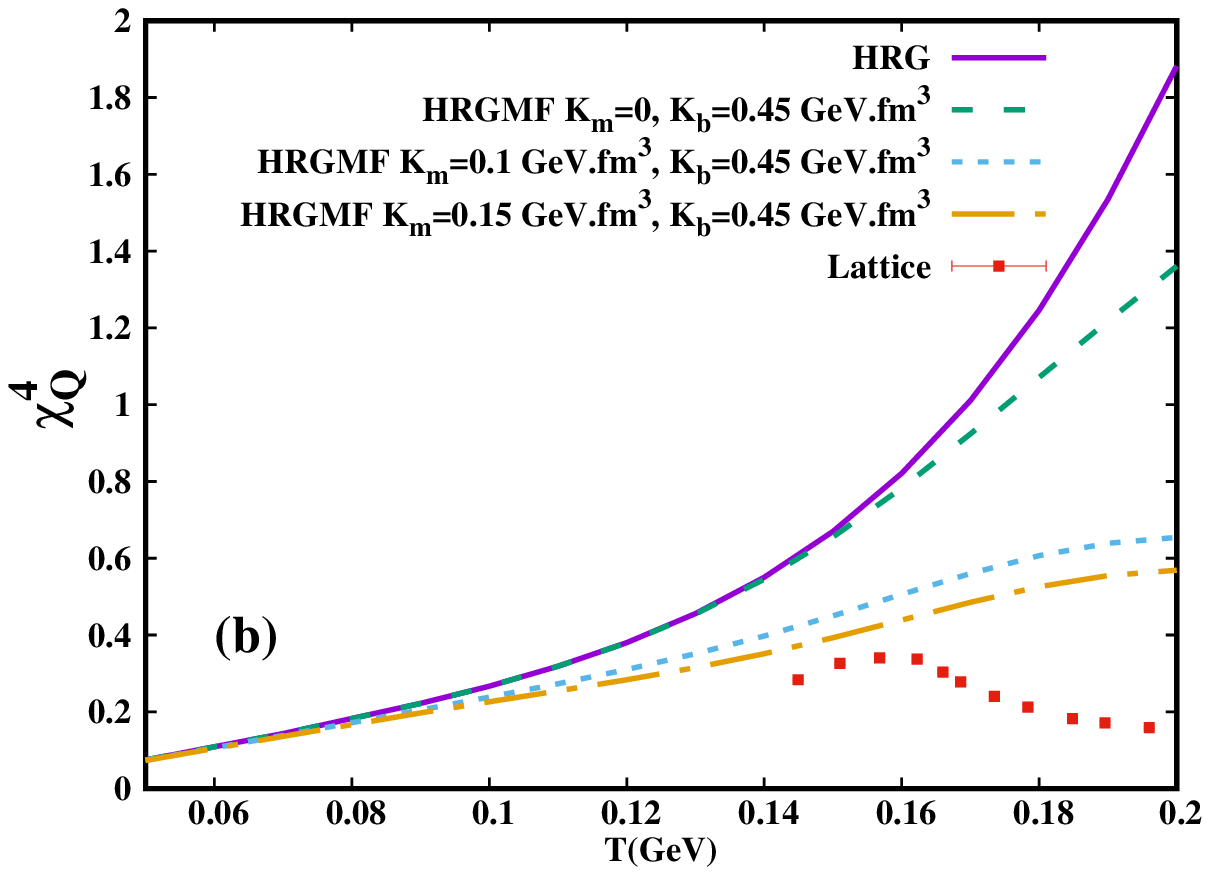}
		\end{tabular}
		\caption{(Color Online) Charge susceptibilities of different orders.   The lattice data is taken from Ref.~\cite{LQCD16}.}
		\label{chiQ}
	\end{center}
\end{figure}
 Fig.\ref{chiQ} shows  2nd and 4th order electric charge susceptibilities. As in the case of baryon number 
susceptibilities, the electric charge susceptibilities increase monotonically with temperature for HRG. However, 
the numerical values of $\chi_Q^2$ and $\chi_Q^4$ are not same for HRG, which is unlike baryon number susceptibilities. 
 We note that there is no qualitative and almost no quantitative difference between results obtained from 
HRG and MFHRG  model when $K_M=0$  except at high temperature. The reason is that the dominant contribution 
to $\chi_Q^{n}$ arises from charged mesons for which there is no repulsive interactions for $K_m=0$. But when we 
switch on mean-field interactions for mesons,  MFHRG model reproduces LQCD results for $\chi_Q^{2}$. Reasonable
 quantitative agreement is achieved for $K_M=0.1$ GeV$\cdot$fm$^{3}$. In case of $\chi_Q^{4}$, 
MFHRG overestimates the susceptibility for all the three choices of $K_M$. 
Nonetheless,  qualitative agreement can be seen for higher values of $K_M$ which also emphasises the 
important role of repulsive interactions among mesons.

\begin{figure}[h]
	\vspace{-0.4cm}
	\begin{center}
		\begin{tabular}{c c}
			\includegraphics[width=8cm,height=8cm]{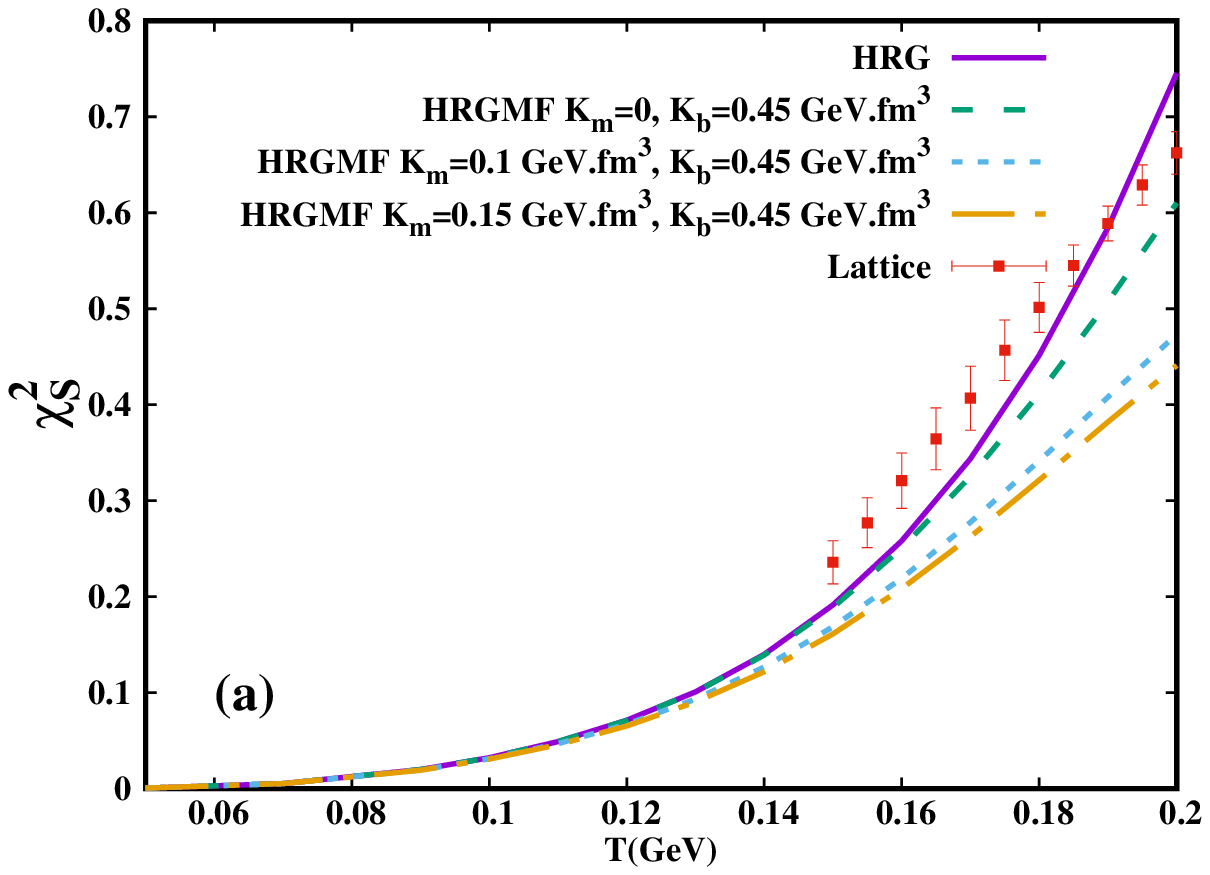} \\
			\includegraphics[width=8cm,height=8cm]{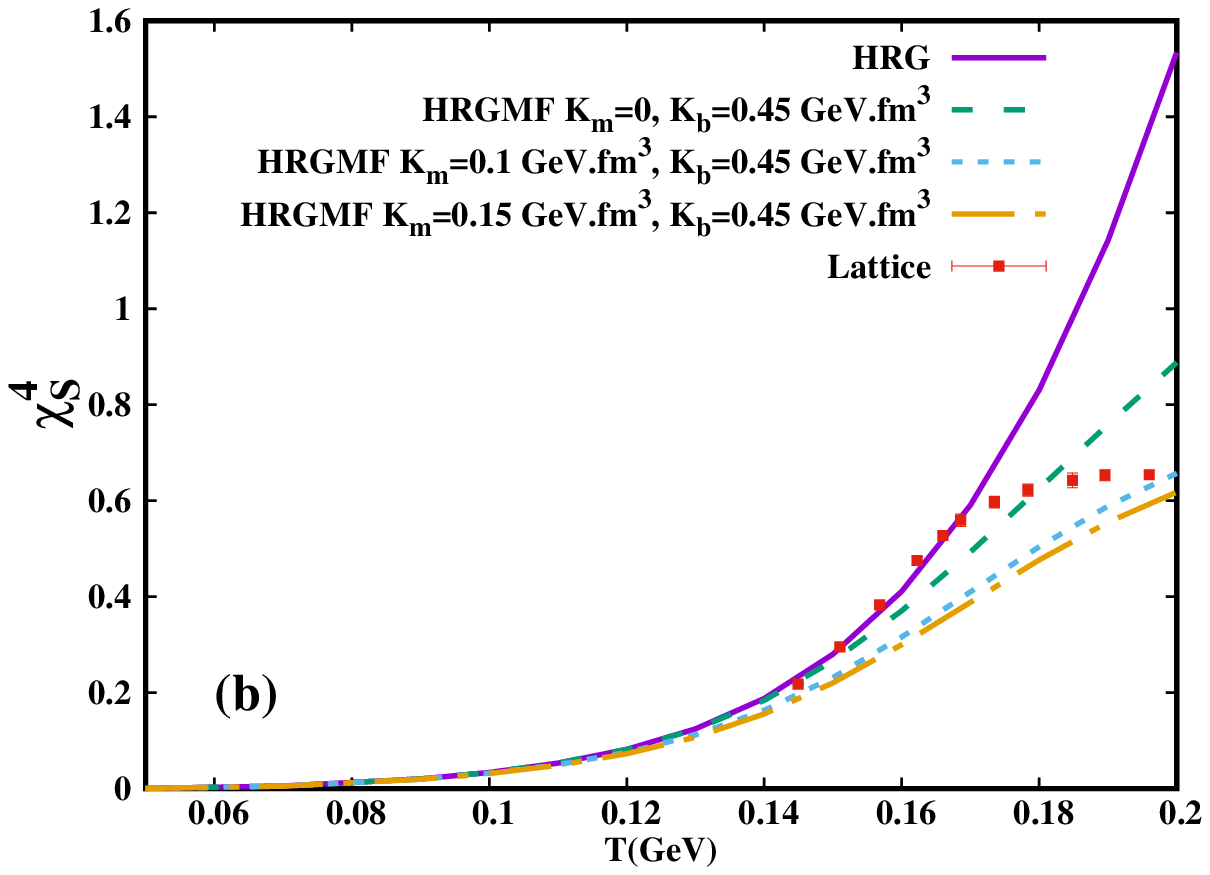}
		\end{tabular}
		\caption{(Color Online) Strangeness susceptibilities of different orders.   The lattice data is taken from Ref.~\cite{LQCD16}. } 
		\label{chiS}
	\end{center}
\end{figure}

In  Fig.\ref{chiS} we have plotted  2nd and 4th order strangeness susceptibilities. We note that the lattice 
results of $\chi_{S}^2$ is well described by ideal HRG model. The results of $\chi_{S}^4$ is 
also reproduced by HRG model up to a temperature of 160 MeV. 
At higher temperatures, the lattice results are closer to the interacting scenario modelled by  
MFHRG model.  We note that the repulsive interactions are found to underestimate strangeness susceptibilities. 
Previous studies have also 
observed similar behaviour of strangeness susceptibilities  as well as correlators 
involving strangeness~\cite{bazavov_strange}.  This observation can be attributed to the unknown strange 
hadronic states not included into the hadronic mass spectrum. In fact, inclusion of these unknown states has 
been found to improve HRG model estimations~\cite{lo}.

 \begin{figure}[h]
	\begin{center}
		\begin{tabular}{c c}
		\hspace{-0.3in}
			\includegraphics[scale=0.65]{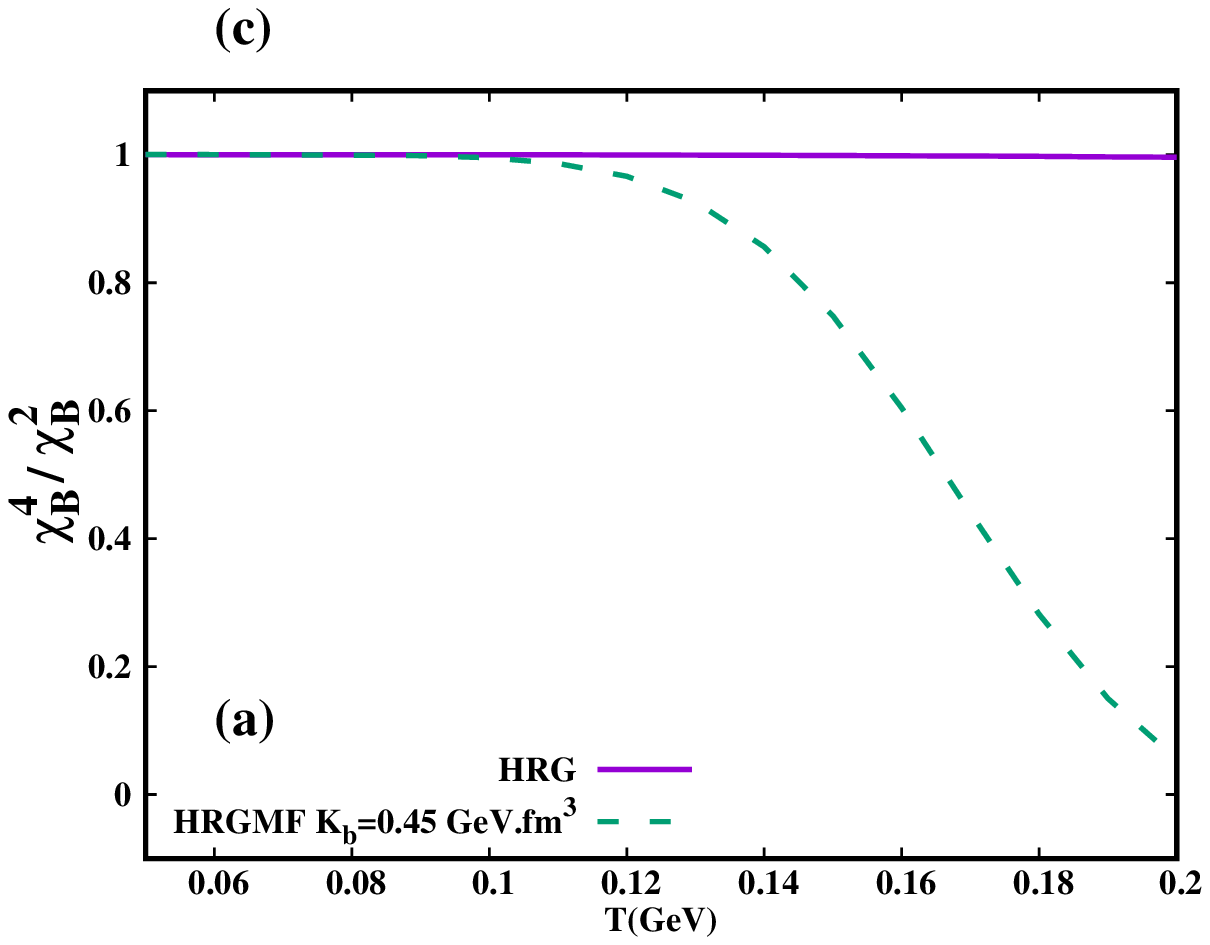}\\
				\hspace{-0.3in}
			\includegraphics[scale=0.65]{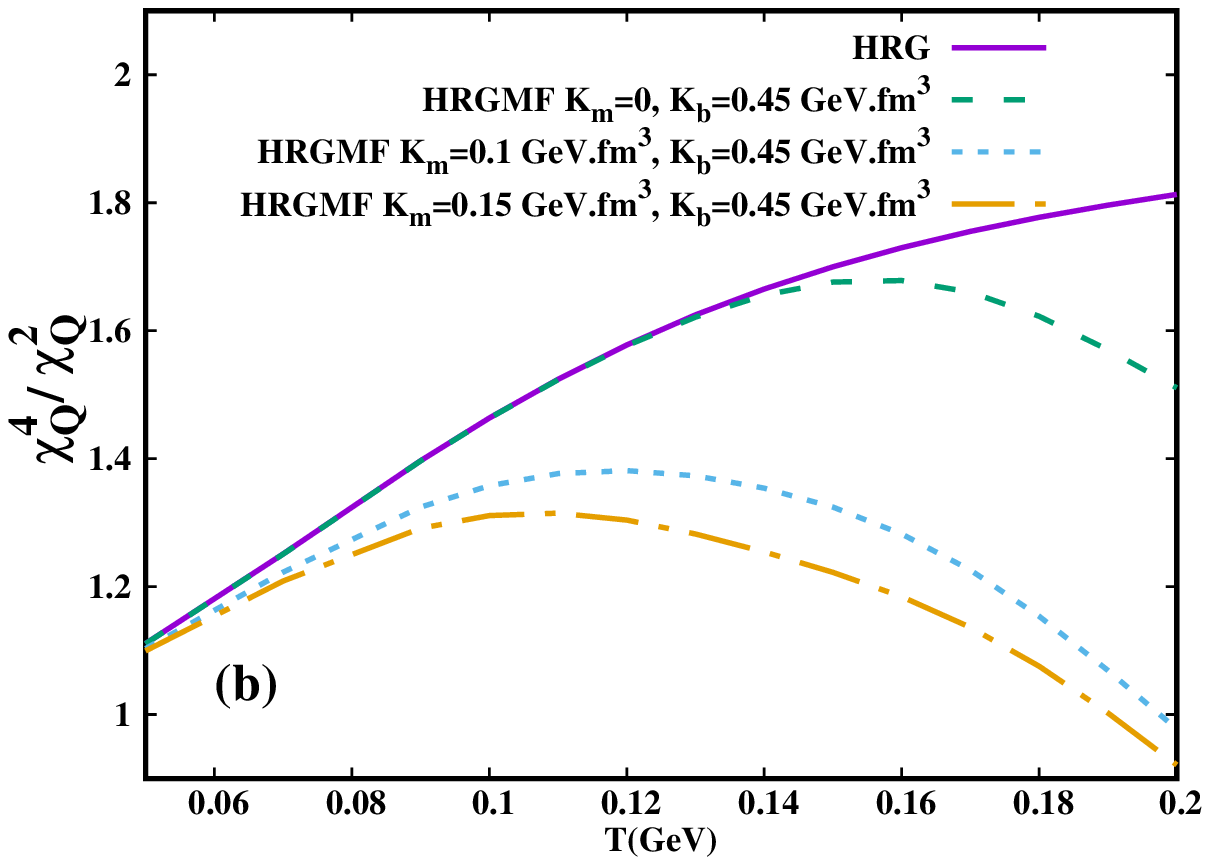}\\
				\hspace{-0.3in}
			\includegraphics[scale=0.65]{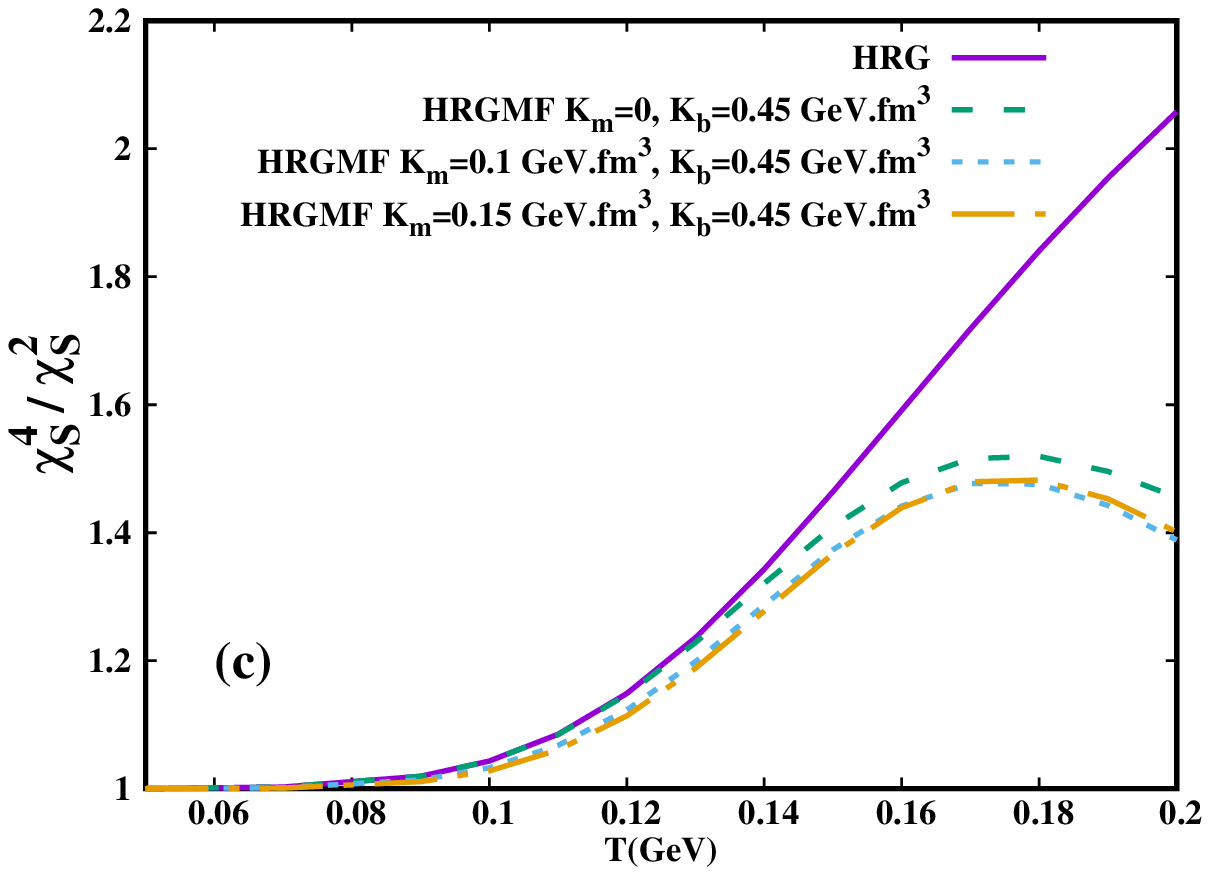}\\
		\end{tabular}
		\caption{(Color Online)  Ratios of fourth order and second order susceptibilities.} 
		\label{chiBQ_ratios}
	\end{center}
\end{figure}

Fig.\ref{chiBQ_ratios} shows ratios of 2nd and 4th order susceptibilities of  baryon, charge and strangeness.
 These ratios are related to the moments of conserved charge fluctuations and hence they are important in 
the context of heavy-ion collision experiments. 
In the MFHRG model,  as mentioned earlier, in the Boltzmann approximation,  (and in the limit of $\beta U\ll 1$),
the susceptibility $\chi_{B}^{n}$ can be approximated as
\begin{equation}
\chi_{B}^{n} =(\chi_{B}^{id})^{n} -2^n\beta^4K_{B}(n_{B}^{id})^2
\label{chiBoltz}
\end{equation}
where $(\chi_{B}^{id})^{n}$ is the n$^{\text{th}}$ order non-interacting susceptibility.
 We note from the above  equation that when repulsive interactions are switched off 
($K_B=0$) one  gets $\frac{\chi_{B}^4}{\chi_B^2}=1$. In HRG, 
and in the Boltzmann approximation, net-baryon kurtosis  show expected Skellam 
behavior~\cite{Asakawa:2015ybt,Vovchenko:2016rkn}. 
The effect of repulsive interactions (2nd term in Eq.(\ref{chiBoltz})) is to 
decrease this ratio. This is because the decrease of the susceptibility compared to ideal HRG
due to the repulsive interaction increase with the order of the susceptibilty.
  This behaviour is also consistent 
with  the lattice QCD data obtained in Ref.~\cite{bazavov_prd95}. 
It was shown in Ref.~\cite{bazavov_prd95} that at low temperature the ratio has 
a value unity and it decreases with temperature to reach the free quark limit at high 
temperature. Thus our result, of MFHRG model,
 reproduces the lattice result. Deviation from the Skellam behaviour can be attributed to 
the repulsive interactions 
between baryons, and again, we cannot neglect its contribution in the conserved charge fluctuation studies. 
 
For electric charge susceptibilities the mesons contribute dominantly and hence Boltzmann approximation will
 not be  good approximation.
Thus, simple expression similar to Eq.(\ref{chiBoltz}) cannot be obtained. Further, in the HRG model only baryons with 
baryon number $B = 1$ contribute to  $\chi_{B}^{n}$, while in case of $\chi_{Q}^n$ multiple charged hadrons contribute. 
In fact, these multiply charged hadrons get larger weight in higher order fluctuations  where both meson as well as 
baryons contribute. Upshot of this is the characteristic deviation of $\frac{\chi_{Q}^4}{\chi_Q^2}$ seen in Fig.\ref{chiBQ_ratios}(b). 
The effect of repulsive interactions is to suppress the number density at high temperature.  This can be seen in 
Fig.\ref{chiBQ_ratios}(b) as the ratio  $\frac{\chi_{Q}^4}{\chi_Q^2}$ in MFHRG deviate from HRG model results. 

The ratio of strangeness susceptibilities more or less gives a similar picture. Like the electric charge scenario,
 in the case of strangeness also the Boltzmann approximation is not valid. Furthermore,
 the particles with multiple strangeness contribute to the scenario. For HRG, the ratio increases monotonically with temperature.
 Once we switch on the interaction the ratio is suppressed. Furthermore,
 the ratio becomes non-monotonic at high temperatures. The repulsive interaction seems to have a significant contribution in this quantity. 

 \begin{figure}[h]
	\begin{center}
		\begin{tabular}{c c}
		\hspace{-0.3in}
			\includegraphics[scale=0.65]{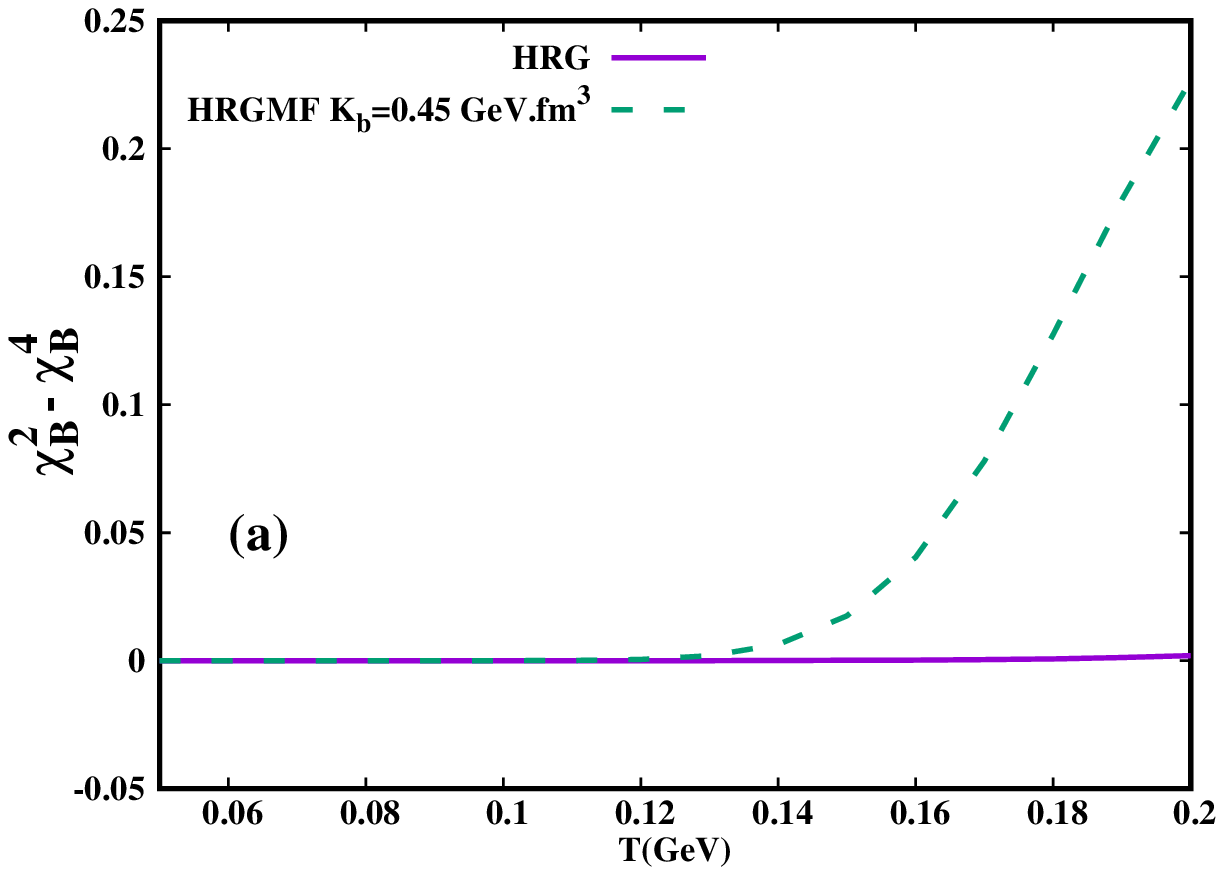} \\
			\hspace{-0.3in}
			\includegraphics[scale=0.65]{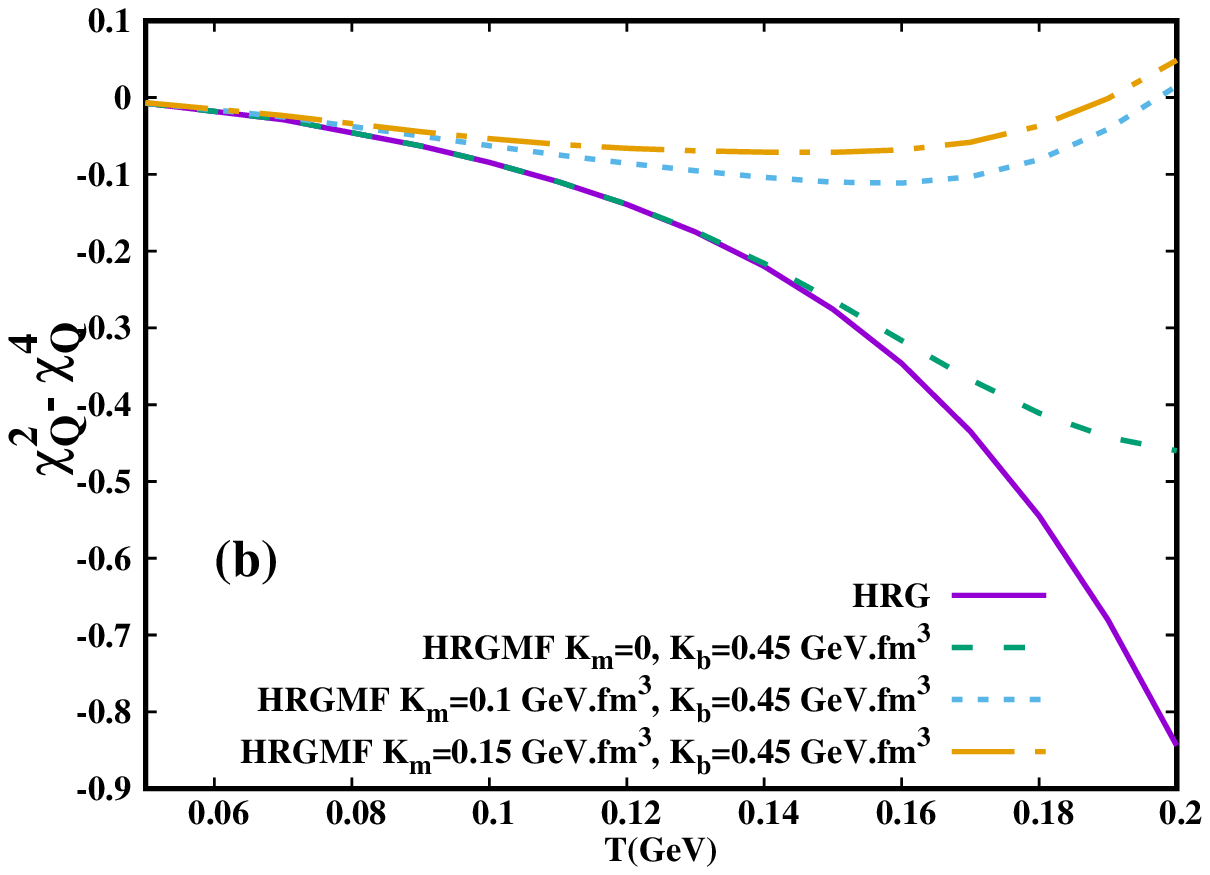} \\
			\hspace{-0.3in}
			\includegraphics[scale=0.65]{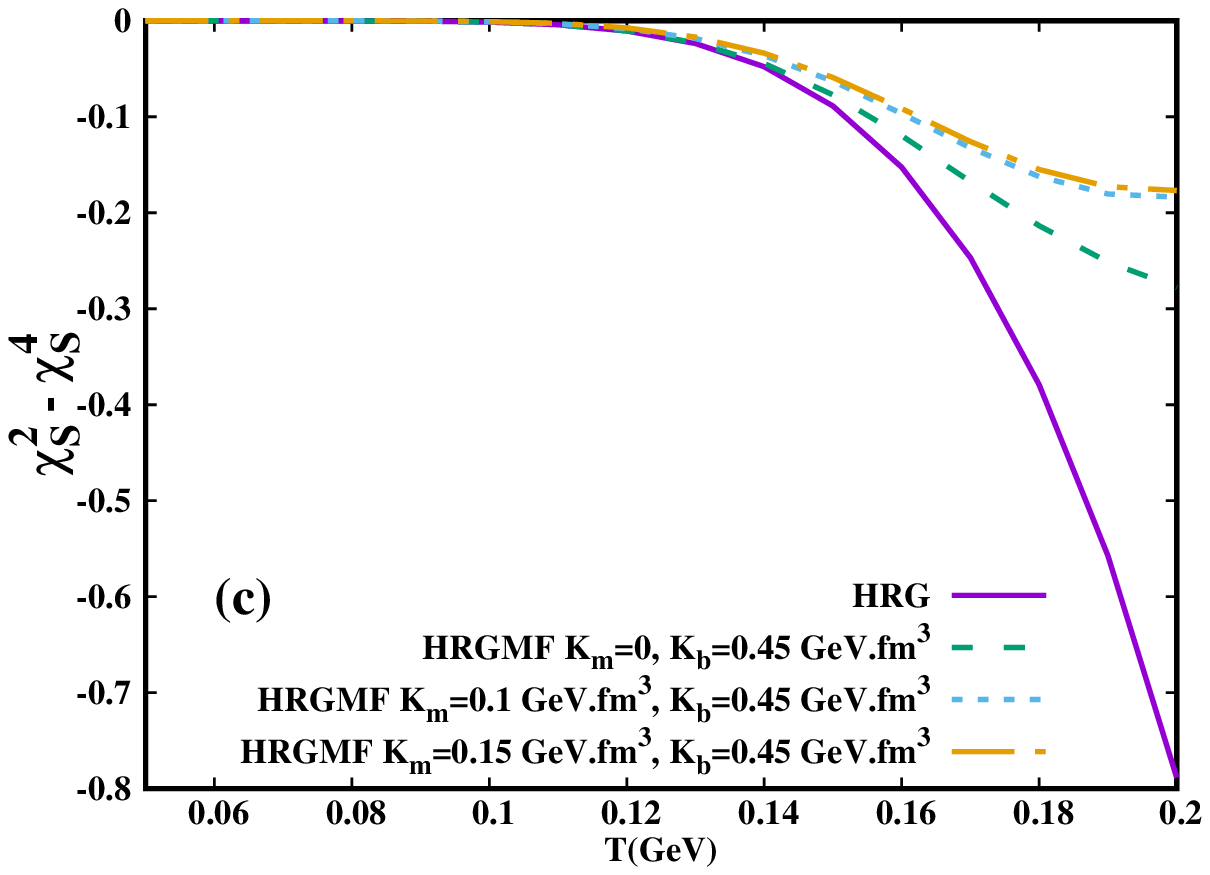} \\
		\end{tabular}
		\caption{(Color Online) Differences of second order and fourth order susceptibilities. } 
		\label{chiBQ_diff}
	\end{center}
\end{figure}
    In case of ratios of susceptibilities it may not be easy to separate the effect of repulsive interactions from medium effects, like in-medium mass modification or the widening of spectral width. If we take the difference of susceptibilities  the results are independent of mass spectrum included in the HRG model.  Fig.\ref{chiBQ_diff} shows differences of 2nd and 4th order susceptibilities. The difference, $\chi^2_B-\chi^4_B$ is zero in HRG model. But if we include the repulsive interactions using mean-field approach then  this difference increases with temperature. This behaviour is in agreement  with LQCD results~\cite{ratti1}. The lattice result shows that the difference increases with temperature as has been found in the MFHRG model.   The charge and strangeness sectors show different behaviour as opposed to baryon sector in HRG model. 
The differences of susceptibilities decrease with increase in temperature. This observation can again be attributed to multiple charged hadrons which contribute more to higher order susceptibilities. The effect of repulsive interactions is to suppress the heavier charged hadrons density at high temperature. Hence we observe less steeper decrease in MFHRG as compared to HRG.
\begin{figure}[t]
	\vspace{-0.4cm}
	\begin{center}
		\hspace{-0.3in}
			\includegraphics[scale=0.65]{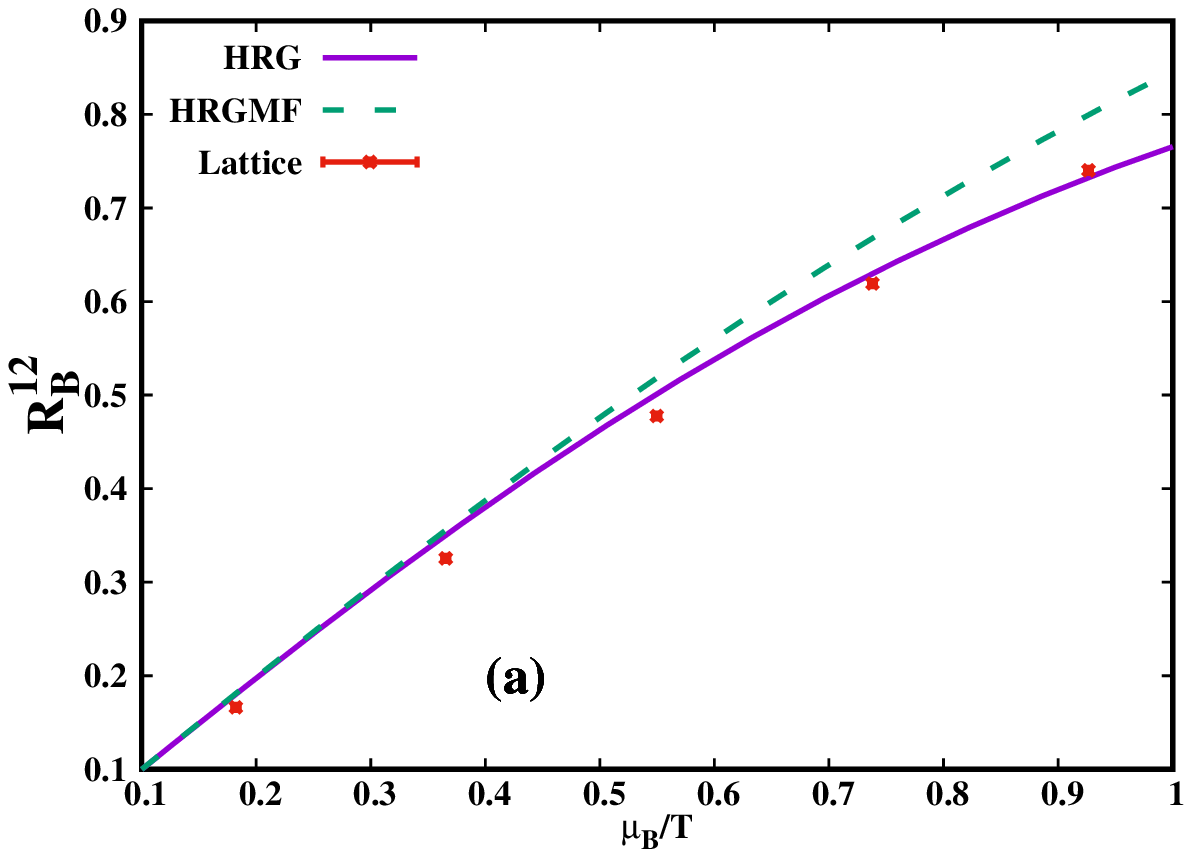} \\
			\hspace{-0.3in}
			\includegraphics[scale=0.65]{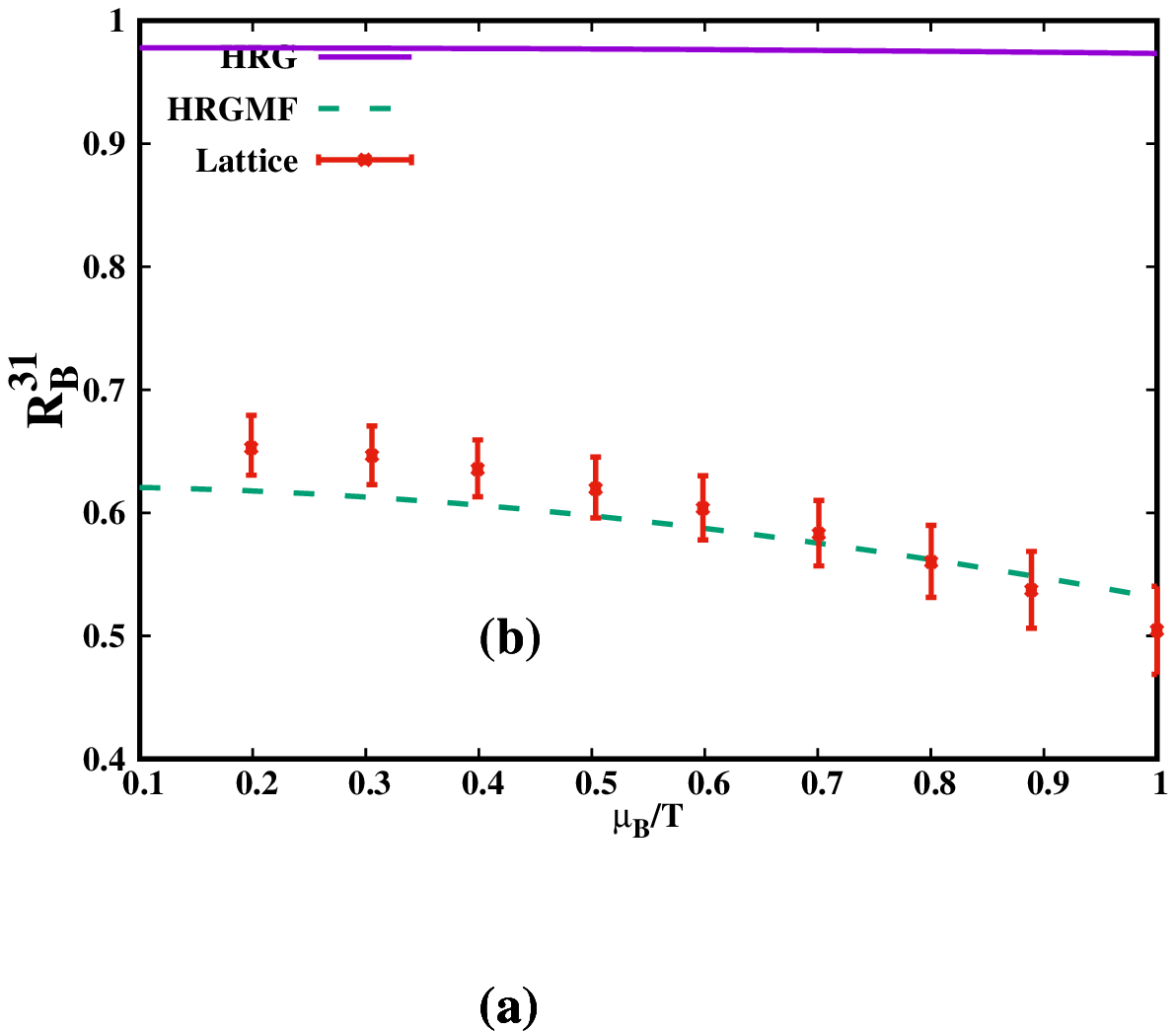} \\
			\hspace{-0.3in}
			\includegraphics[scale=0.65]{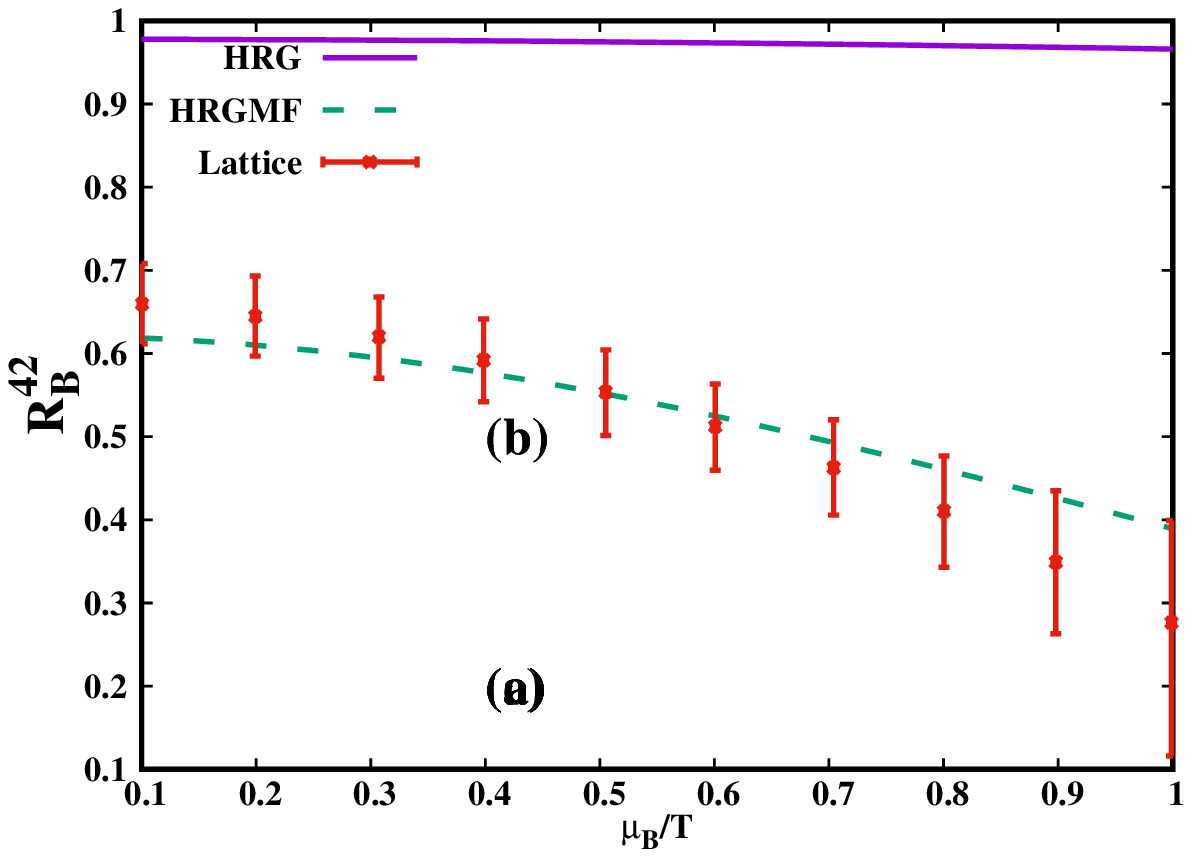}
\caption{(Color Online) $R_B^{12}$, $R_B^{31}$ and $R_B^{42}$ (defined in the text) as 
functions of $\mu_B/T$. The lattice data is taken from Ref.~\cite{LQCD20}. } 
		\label{rb1}
	\end{center}
\end{figure}

So far, we have shown the results for vanishing baryon chemical potential. Next, we discuss the case of non-vanishing
values for the chemical potentials $\mu_B$, $\mu_Q$ and $\mu_S$. As shown in Ref.~\cite{LQCD20}, in general, the susceptibilities
at finite chemical potential can be expanded in powers of $\mu_i/T$ $(i=B,Q,S)$ with coefficients 
being generalised susceptibilities
that can be evaluated at vanishing chemical potentials. The constraints of fixed strangeness and electric charge to 
baryon number fixes the relation between the electric charge chemical potential and the strangeness chemical potential 
to the baryon chemical potential at a given temperature~\cite{LQCD17}. As in Ref.~\cite{LQCD20}, we shall consider the case of strangeness neutral system
i.e. $n_S=0$ and $n_Q/n_B=0.4$ which are representative conditions met in heavy ion collision experiments with gold or uranium 
nuclei. Thus 
one has the relation with $\hat\mu_i=\mu_i/T$
$$\hat \mu_Q\simeq q_1(T)\hat\mu_B+q_3(T)\hat\mu_B^3$$
\be
\hat \mu_S\simeq s_1(T)\hat\mu_B+s_3(T)\hat\mu_B^3
\label{muqs}
\ee

The coefficient functions $q_i$ and $s_i$ are given in terms of the susceptibilities at zero chemical potential and the
ratio $r=n_Q/n_B$ as in Ref.~\cite{LQCD172}. We have calculated here, these susceptibilities in MFHHRG model to calculate the
coefficients on RHS of Eq.(\ref{muqs}). Different susceptibilities are then estimated using Eq.(\ref{chin}) for  a given
$\mu_B$, $\mu_Q$ and $\mu_S$.

In Fig.\ref{rb1}, we have plotted the ratio of the cumulants for the net baryon number fluctuations as functions 
of $\mu_B/T$. We have taken here, $T=158$ MeV as in Ref.~\cite{LQCD20} corresponding to 
the upper end of the error band of the pseudocritical temperature for vanishing chemical potential.
On the top panel
we have plotted $R^{12}_B$, the ratio of mean to variance of net baryon number i.e $\chi^1_B(T,\mu_B)/\chi^2_B(T,\mu_B)$ 
as a function of $\mu_B/T$. On the middle panel, we have plotted the ratio of skewness ratio $R^{31}_B(T,\mu_B)=
\frac{\chi^3_B(T,\mu_B)}{\chi^1_B(T,\mu_B)}$. Similarly, in the bottom panel, we have plotted the kurtosis 
ratio i.e. $R^{42}_B=\frac{\chi^4_B(T,\mu_B)}{\chi^2_B(T,\mu_B)}$ as a function of $\mu_B/T$. 
As may be observed, the lower cumulant i.e. $R^{12}_B$ from lattice simulations
are in good agreement with the HRG model estimation which takes the hadrons to be point like and noninteracting. 
However, the simple HRG model is inadequate to describe the behaviour of  the higher order cumulants 
$R_{31}^B$ and $R_{42}^B$ obtained from lattice simulations. In fact, for simple HRG model, the values of both 
these higher order cumulant ratios are unity. 
When one includes the repulsive interactions among hadrons within MFHRG model, 
the lattice QCD results seem to be in agreement with the MFHRG model even at finite chemical potential. 
We have taken here the value of the baryonic repulsive
parameter $K_B=0.45$ GeV$\cdot$fm$^{3}$ as was taken for zero baryon density case. Further, 
the baryonic susceptibilities are independent of $K_M$.

  \section{summary}
 \label{secIV}
We have studied  the effect of repulsive interaction on the susceptibilities of conserved charges. For this purpose we have used the MFHRG model. We have two parameters, namely $K_B$ and $K_M$, to describe the baryonic and mesonic interactions respectively. We have kept $K_B$ fixed at $0.45$ GeV$\cdot$fm$^{3}$ and varied $K_m$ in the range $0-0.15$ GeV$\cdot$fm$^{3}$. Our results have been confronted with the lattice data. We conclude the following important points :

\begin{itemize}

\item The results of HRG model, as such, cannot explain the LQCD data for baryonic susceptibilities. 
A repulsive interaction, through MFHRG model, is required to explain the data for the higher order susceptibilities. The baryonic susceptibilities are independent of mesonic interactions. \\

\item The repulsive interactions are further required to reproduce charge susceptibilities. The $\chi_Q^2$ is better reproduced with interactions compared to HRG. However $\chi_Q^4$ is over estimated with the choice of the $K_m$ that we use though there is a qualitative matching. The 
trend shows a stronger mesonic interaction may reproduce data  better. \\

\item The fourth order strangeness susceptibility is better reproduced, with the repulsive interactions, 
as compared to the second order susceptibility. However, calculation of strangeness susceptibilities have some inherent problems. The problem arises due to unknown strange hadronic states not included into the hadronic mass spectrum. \\

\item The ratios and differences of susceptibilities provide important information regarding interactions. These quantities provide us important  details regarding the deviation from ideal gas scenario. Both the ratios and differences of susceptibilities are in conformity with the corresponding available lattice data. \\

\item The repulsive interactions for baryons become particularly important for finite density 
results. Our results, for the ratios of  susceptibilities at finite density, have an excellent match 
with  those obtained from the LQCD simulations. 

\end{itemize}

\section{Acknowledgement}

G. K. is financially supported by DST-INSPIRE Faculty research grant number DST/INSPIRE/04/2017/002293.  S.P. is financially supported by UGC, New Delhi.

\end{document}